\documentclass[sigconf,nonacm ]{acmart} 
\AtBeginDocument{%
  }

\usepackage[table]{xcolor}
\usepackage{enumitem}
\usepackage{makecell}
\usepackage{graphicx} 

\begin{document}

\title{From Forecasts to Auditable Reports: Evidence Contracts for LLM-Assisted Housing-Guarantee Risk Monitoring}

\author{Hyeongcheol Kim}
\affiliation{%
  \institution{Pusan National University}
  \institution{Korea Housing \& Urban Guarantee Corporation}
  \country{Republic of Korea}
}
\email{polpokpoi@khug.or.kr}

\author{Yoontae Hwang}
\authornote{Corresponding author.}
\affiliation{%
  \institution{Pusan National University}
  \country{Republic of Korea}
}
\email{yoontae.hwang@pusan.ac.kr}

\begin{abstract}
Translating next-month housing-guarantee risk forecasts into auditable operational reports is essential yet challenging because upper-tail events are sparse, source records are confidential, and generated narratives can distort the underlying evidence. Using monthly South Korean \textit{jeonse} deposit guarantee data from September 2015 to December 2025, we introduce an evidence-constrained reporting pipeline that prioritizes upper-tail monitoring, retrieves historical precedents aligned with the forecasting rationale, organizes admissible information into typed evidence contracts, and verifies generated claims before analyst review. We train and select the forecasting backbone on the original panel, whereas the reporting experiments use synthetic aggregate scenarios calibrated to its empirical ranges and temporal structure. The selected forecasting model substantially improves high-risk detection while retaining competitive average error. Across eight LLMs, structured evidence consistently increases report quality, numerical fidelity, and claim-level grounding. A practitioner evaluation involving 51 analysts and related domain professionals further indicates that the reports support real-world review and decision-making: most participants rated them as practically useful and endorsed an operational pilot. These findings demonstrate that reliable LLM-assisted reporting requires predictive models to be coupled with structured evidence, explicit verification, and analyst oversight.
\end{abstract}

\maketitle

\section{Introduction}
Public guarantee institutions must convert noisy monthly forecasts into warnings that are both timely and auditable. This task is particularly consequential in housing-guarantee markets, where a missed warning can delay liquidity planning, slow case supervision, and weaken preparedness for cascading guarantee defaults. False warnings are also costly because they consume analyst capacity on segments that may not require intervention. {At the same time, growing use of AI assistance raises a governance requirement: warnings must be evidence-verifiable without exposing raw records or allowing a language model to invent support. The central challenge is therefore evidence-constrained risk communication, not prediction alone.}

\begin{figure}[ht!]
\begin{center}
\includegraphics[width=1.0\linewidth]{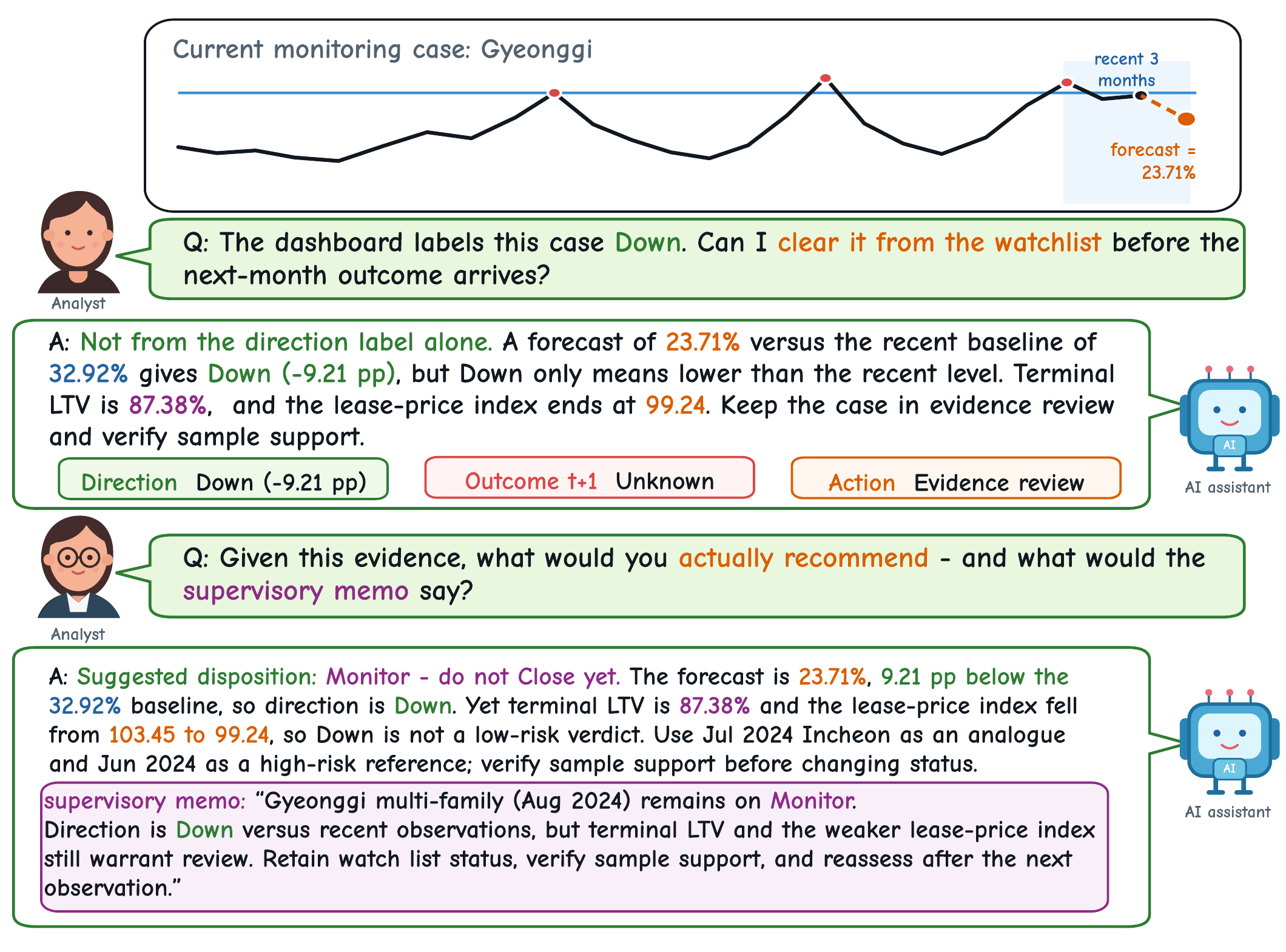}
\end{center}
\vspace{-3mm}
\caption{Why \textit{Down} does not imply \textit{Close}. The 23.71\% forecast is 9.21 percentage points below the recent baseline, yet the 87.38\% terminal loan-to-value ratio and declining lease-price index support continued review. Directional labels therefore require contextual evidence and analyst judgment.}
\vspace{-5mm}
\label{Figure_1}
\end{figure}

We examine this challenge in South Korea's \textit{jeonse} deposit guarantee market \citep{ambrose2003modeling, moon2018housing}. Under a \textit{jeonse} contract, a tenant pays a large upfront deposit that should be returned at maturity, while a public institution guarantees repayment when the landlord defaults. Risk reflects macroeconomic conditions, local housing prices, collateral burdens, and regional demand, which interact over different timescales and vary across region-by-housing-type segments. Monitoring is therefore difficult because severe events are sparse and market signals often arrive with a lag. In practice, analysts require more than a point forecast: they need an interpretable evidence trail that indicates which segments warrant attention and why.

Direct forecasting and naive report automation are insufficient for three reasons. First, monitoring agencies face asymmetric costs: under-predicting right-tail events can impair operations more severely than over-predicting low-risk cases \citep{gneiting2011quantiles, gneiting2011making}. A model that minimizes average error is therefore not necessarily suitable for a risk desk. Second, data-governance requirements restrict the transmission of raw guarantee records to external large language models (LLMs) \citep{carlini2019secret, carlini2021extracting}, complicating automated narrative generation under data-minimization constraints. Third, accountability requires an analyst-in-the-loop workflow \citep{amershi2019guidelines}. Linguistic fluency cannot offset operational failures when a model fabricates trend directions, numerical units, reference periods, or historical analogues \citep{ji2023survey}.

To address these limitations, we formulate monitoring as a process of evidence construction. The forecast is the first component rather than the final output. Each warning is paired with model-derived diagnostic signals, including emphasized variables, salient lookback periods, comparable historical windows, and a bounded evidence record that specifies the admissible claims. We treat these signals as audit-oriented evidence rather than causal explanations. The language model neither estimates risk nor retrieves additional facts. It converts the evidence record into a draft whose factual claims can be checked automatically and reviewed by an analyst. This formulation transforms open-ended generation into bounded, verifiable reporting.

Our implementation combines an interpretable time-series forecaster with representation-aligned retrieval. A Temporal Fusion Transformer \citep{lim2021tft} predicts next-month segment risk and exposes variable-selection and temporal-attention signals. We combine these signals into a window-level rationale representation and use Centered Kernel Alignment to retrieve historical windows that are similar in the model's reasoning space \citep{cortes2012cka, kornblith2019cka}. Rather than matching cases solely by raw values, the retrieval step asks whether the current warning and a historical precedent are supported by comparable variable-time patterns. The resulting precedents and forecast evidence are then organized into a structured report interface. Our experiments evaluate this pipeline by jointly considering average forecast error, upper-tail monitoring utility, and the fidelity of the evidence used in reporting.

Our contributions are threefold:
\begin{itemize}
    \item We formulate housing-guarantee monitoring as the conversion of sparse risk forecasts into auditable review evidence under capacity and governance constraints.
    \item We develop an evidence-construction workflow that links forecast values, model-derived drivers, influential periods, and historical precedents before report generation.
    \item We evaluate the workflow through tail-sensitive forecasting utility, report fidelity, auditability, and practitioner assessment.
\end{itemize}

\section{Related Works} \label{sec_related}

\paragraph{Housing-Finance Risk}
Housing-finance risk is inherently a delayed monitoring problem because stress can accumulate before defaults or guarantee accidents become observable. Mortgage-default research links repayment distress to {interest-rate exposure, house-price movements, negative equity, and borrower heterogeneity} \citep{deng2000mortgage}. {Structural models in this research further formalize these links to repayment distress} \citep{campbell2015model}. Evidence from the mortgage crisis shows that leverage and declining local prices amplify default risk \citep{mayer2009rise}, while credit expansion can intensify household vulnerability \citep{mian2009consequences}. Although jeonse guarantees differ from mortgages in contract structure, both require balance-sheet pressure and collateral deterioration to be detected before losses materialize. This motivates a monthly panel of collateral burdens, regional price indices, guarantee exposure, and macro-financial covariates \citep{hwang2023identifying, kim2023household}. It also requires an evaluation criterion aligned with monitoring: a guarantee desk must allocate review capacity while reducing missed high-risk episodes, not merely minimize average error. Asymmetric-loss theory shows that decision-aligned forecasts can differ from {conditional means, or expected average outcomes,} when over- and under-prediction have unequal costs \citep{christoffersen1997optimal}, and flexible-loss evaluation reaches the same conclusion \citep{elliott2005estimation}. We apply this principle in a restricted operational form. {Because the regret objective imposes larger penalties for missed high-risk episodes, model evaluation places greater emphasis on recalling high-risk cases than on average error.} The forecast remains a statistical signal for analyst review rather than an automatic decision.

\paragraph{Interpretable time-series forecasting}
The forecasting model must capture temporal dependence while exposing evidence that can be inspected. LSTMs provide a recurrent benchmark for sequential dependence \citep{hochreiter1997lstm}, DLinear offers a simple linear time-series baseline \citep{zeng2023dlinear}, and LightGBM remains effective for heterogeneous tabular inputs \citep{ke2017lightgbm}. These models are informative comparators, but they do not directly produce the variable-time rationale required for retrieval. Temporal Fusion Transformers are well suited to this role because they combine multi-horizon forecasting with variable selection and temporal attention \citep{lim2021tft}, building on the broader transformer mechanism \citep{vaswani2017attention}. We do not interpret these internal summaries as causal effects \citep{jain2019attention, wiegreffe2019attention}. Interpretability must be defined by the decision context and validated against its downstream use \citep{doshi2017rigorous}, particularly in high-stakes settings where post-hoc explanations can create misplaced confidence \citep{rudin2019stop}. We therefore convert the model's variable-selection and attention outputs into a forecast-rationale matrix that records which variables and lookback periods shaped each prediction and serves as the retrieval representation. Post-hoc methods provide an alternative: LIME estimates local feature importance \citep{ribeiro2016lime}, SHAP unifies attribution methods \citep{lundberg2017shap}, TimeSHAP extends perturbation-based explanations to recurrent sequences \citep{bento2021timeshap}, and Dynamask uses dynamic masks for time-series explanations \citep{crabbe2021dynamask}. Our objective is narrower: rather than adding a stand-alone saliency map, we use the forecaster's rationale matrix as a searchable object that connects prediction to historical evidence.
\vspace{-2mm}
\paragraph{Evidence-aligned retrieval and reporting}
An explanation becomes operationally useful when it points to evidence that analysts can inspect. Raw-covariate matching may overlook cases in which the model relies on similar variable-time patterns. Centered Kernel Alignment compares representational geometry \citep{cortes2012cka} and has been used to assess similarity across neural representations \citep{kornblith2019cka}. We apply CKA to TFT rationale matrices to retrieve leakage-free historical windows with comparable model emphasis. These windows are not causal matches. They are auditable precedents whose relevance can be examined through the same artifacts that support the forecast. The reporting layer must preserve this evidentiary structure. Retrieval-augmented generation conditions language models on external evidence \citep{lewis2020rag}, and large language models can produce fluent task-conditioned text \citep{brown2020language}. However, they can also generate unsupported claims \citep{ji2023survey}. Moreover, automatic evaluation depends on metric and evaluator design \citep{liu2023geval}, RAG-specific assessments face related faithfulness and grounding challenges \citep{es2024ragas}, and LLM evaluators may exhibit fairness or preference biases \citep{wang2024fair}. We therefore insert a structured evidence card between retrieval and generation. The card fixes the forecast, direction label, key drivers, important periods, comparable windows, and caveats before drafting, after which rule-based audits verify that numbers, labels, periods, and features are preserved. This design integrates asymmetric monitoring, interpretable forecasting, representation-aligned retrieval, and bounded report generation into a single auditable workflow.

\vspace{-2mm}
\section{Methodology} \label{method}

\begin{figure*}[ht]
\small{
  \centering
  \includegraphics[width=\linewidth]{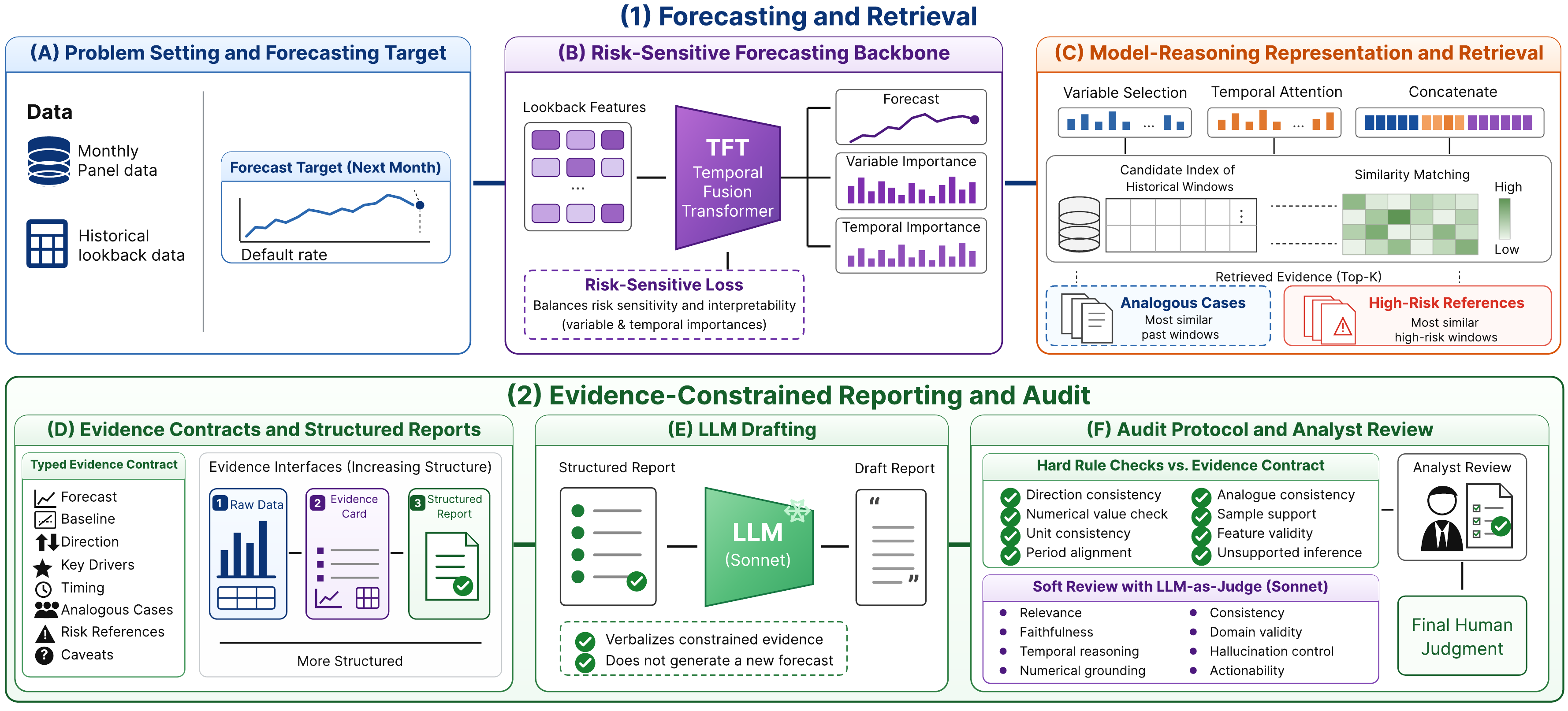}}
  \caption{Overview of the forecast-to-report pipeline. (A,B) A risk-sensitive Temporal Fusion Transformer forecasts next-month segment risk and exposes model-derived signals. (C) Centered Kernel Alignment retrieves admissible historical windows with similar rationales while separating analogous cases from high-risk references. (D,E) Typed evidence contracts and structured interfaces constrain the LLM to supported claims. (F) Deterministic checks audit factual fidelity, rubric scoring supports offline evaluation, and final judgment remains with analysts.}
  \label{figure_2}
 \vspace{-12pt}
\end{figure*}

\paragraph{Problem Setting}
\label{subsec_problem_setting}

Let $i$ index a region-by-housing-type segment and $t$ a monthly guarantee-completion period. The observed outcome is the segment-level guarantee accident rate $y_{i,t}$, defined as the number of guarantee accidents divided by the number of completed guarantees in that segment and month. The information set available before month $t+1$ is denoted by $\mathcal{F}_{i,t}$ and contains a length $L$ history of guarantee exposure, collateral burden, housing-market indicators, and macro-financial covariates. The forecasting task estimates the next-month rate as $\widehat y_{i,t+1}=f_{\theta}(\mathcal{F}_{i,t})$.

Report generation begins only after the forecast is fixed. Given $\widehat y_{i,t+1}$, a recent baseline $b_{i,t}$, model-derived interpretation signals, retrieved historical windows, and an evidence contract $\mathcal{E}_{i,t}$, the system produces a draft report $r_{i,t}$. A valid draft must satisfy two constraints: it cannot introduce a new risk estimate, and every numerical, temporal, feature, or analogue claim must be supported by the evidence contract. This separation restricts the LLM to conditional text generation rather than statistical estimation.

Directional reporting compares the forecast with a recent moving-average baseline. Let $b_{i,t}$ denote this baseline and $\tau$ a pre-specified threshold measured in\%p units. The direction label is defined by
\begin{equation}
D_{i,t+1}
= \begin{cases} \mathrm{Up} & \text{if } \widehat y_{i,t+1}-b_{i,t} \geq \tau,\\
                \mathrm{Down} & \text{if } \widehat y_{i,t+1}-b_{i,t} \leq -\tau,\\
                \mathrm{Flat} & \text{otherwise.}
\end{cases}
\label{eq_direction_rule}
\end{equation}
Because the threshold is fixed before generation, the rule is a deterministic report field rather than an estimate of economic loss. Its output can therefore be verified before analyst review.

\paragraph{Risk-Sensitive Forecasting Backbone}
\label{subsec_forecasting_backbone}

The forecasting module must support both prediction and evidence construction. A scalar forecast identifies segments that may require attention, while interpretation signals indicate why a segment was surfaced and provide keys for historical retrieval. Temporal Fusion Transformers are suitable because their variable-selection networks and temporal-attention layers expose feature- and period-level signals in the same forward pass that generates the forecast \citep{lim2021tft}. Conventional average-error losses aid calibration but do not represent the monitoring asymmetry faced by a risk desk, for which under-predicting a right-tail episode may be more consequential than over-predicting a low-risk segment. We therefore adopt a regret-style objective that places additional weight on under-prediction while retaining the interpretation outputs required downstream \citep{elliott2005estimation}. The core loss is
\begin{equation}
\mathcal{L}_{\mathrm{reg}} (\widehat y_{i,t+1},y_{i,t+1})
= c^{\mathrm{over}}_{i,t+1} (\widehat y_{i,t+1}-y_{i,t+1})_{+}
+ \lambda c^{\mathrm{under}}_{i,t+1} (y_{i,t+1}-\widehat y_{i,t+1})_{+}.
\label{eq_regret_loss}
\end{equation}
Here $(a)_{+}=\max(a,0)$. The terms $c^{\mathrm{over}}_{i,t+1}$ and $c^{\mathrm{under}}_{i,t+1}$ are pre-specified monitoring weights, and $\lambda$ controls the relative under-prediction penalty. This objective is not intended to reproduce the institution's complete funding, capital, or provisioning formula. Its narrower role is to shift the operating point toward high-risk visibility while preserving model-derived evidence for the reporting layer.

\paragraph{Model-Reasoning Representation and Retrieval} \label{subsec_reasoning_retrieval}
We next construct a representation that makes model evidence comparable across time. Raw time-series proximity is inadequate because windows with similar realized rates may reflect different collateral or macro-financial conditions, whereas windows with different outcomes may induce similar emphasis over variables and lookback periods. Accordingly, the retrieval layer compares historical windows in the forecaster's reasoning space rather than in the raw covariate space.

For segment $i$ at monitoring date $t$, the TFT produces variable-selection and temporal-attention weights from the same lookback window used for prediction. Let $M^{\mathrm{VSN}}_{i,t}\in\mathbb{R}^{L\times m}$ denote the variable-selection matrix over $m$ inputs, and let $M^{\mathrm{Attn}}_{i,t}\in\mathbb{R}^{L\times H}$ denote the temporal-attention matrix over $H$ retained heads. We define the window-level model-reasoning representation as
\begin{equation}
Z_{i,t}= [ M^{\mathrm{VSN}}_{i,t} \,\Vert\, M^{\mathrm{Attn}}_{i,t} ] \in \mathbb{R}^{L\times(m+H)}.
\label{eq_interpretation_matrix}
\end{equation}
This representation is not a causal decomposition of guarantee risk. It records how the trained forecaster distributes importance across observed variables and historical periods. Feature and temporal evidence answer distinct interpretive questions \citep{kim2024cafo}. In our workflow, this distinction becomes operational because $Z_{i,t}$ functions as a retrieval key rather than a stand-alone explanation.

At monitoring date $t$, the candidate pool is restricted to historical observations that would have been available at that date. Each candidate window $(j,s)$ has a reasoning matrix $Z_{j,s}$ and metadata describing its segment, anchor month, target month, realized accident rate, exposure size, and accident count. This restriction prevents future information from entering the retrospective workflow and makes the retrieved cases admissible as analyst-facing evidence. For target window $(i,t)$, we compute centered linear CKA against every admissible candidate. Let $\bar Z$ denote the matrix obtained by removing column means. The similarity score is
\begin{equation}
\mathrm{CKA}(Z_{i,t},Z_{j,s})=
\frac{\lVert \bar Z_{i,t}^{\top}\bar Z_{j,s} \rVert_F^2}
     {\lVert \bar Z_{i,t}^{\top}\bar Z_{i,t} \rVert_F \lVert \bar Z_{j,s}^{\top}\bar Z_{j,s} \rVert_F}.
\label{eq_cka}
\end{equation}
CKA compares representational geometry rather than a single flattened distance. Because it is invariant to orthogonal transformations and isotropic rescaling, it is appropriate for comparing model-derived evidence matrices whose columns combine feature- and time-level signals \citep{cortes2012cka, kornblith2019cka}. The highest-scoring candidates are returned as analogous cases. They contextualize the forecast but neither update it nor determine the risk threshold. High-risk reference windows are selected separately according to realized risk and exposure support. This distinction is essential: a severe historical outcome may be informative without resembling the model's rationale for the current window. Separating the two evidence types prevents the LLM from treating high realized risk as proof of reasoning similarity.

\paragraph{Evidence Contracts and Structured Report}
\label{subsec_evidence_contracts}

Retrieved objects become useful for generation only after they are converted into a form that an LLM can cite without changing their meaning. The CKA layer returns analogous windows, high-risk reference episodes, raw target histories, and sample-support information, but these objects do not by themselves constrain which number, period, driver, or precedent the model may emphasize. {The evidence contract fixes these choices through deterministic modules before the LLM is invoked.} For each target window, the forecasting and retrieval components produce the following typed fields:
\begin{equation}
\vspace{-2pt}
\mathcal{E}_{i,t} =
\{
e^{\mathrm{fcst}}_{i,t},
e^{\mathrm{base}}_{i,t},
e^{\mathrm{dir}}_{i,t},
e^{\mathrm{drv}}_{i,t},
e^{\mathrm{time}}_{i,t},
e^{\mathrm{sim}}_{i,t},
e^{\mathrm{risk}}_{i,t},
e^{\mathrm{cav}}_{i,t}
\}.
\label{eq_evidence_contract_simple}
\vspace{-1.5pt}
\end{equation}
The field $e^{\mathrm{fcst}}_{i,t}$ stores the forecast, $e^{\mathrm{base}}_{i,t}$ the recent baseline, and $e^{\mathrm{dir}}_{i,t}$ the direction label from Eq.~\ref{eq_direction_rule}. The fields $e^{\mathrm{drv}}_{i,t}$ and $e^{\mathrm{time}}_{i,t}$ contain the selected variable and lookback evidence; $e^{\mathrm{sim}}_{i,t}$ stores CKA-based analogues; $e^{\mathrm{risk}}_{i,t}$ stores high-risk references selected by realized outcomes and exposure support; and $e^{\mathrm{cav}}_{i,t}$ records sample support, exposure, observed accident counts, and small-cell warnings. Every field is computed before the LLM is invoked.

The evidence contract has a deliberately narrow role. It is neither a second forecasting model nor a free-form explanation, but a typed set of admissible report facts. In compact notation, $\mathcal{E}_{i,t}=g_{\mathrm{EC}}(\widehat y_{i,t+1}, b_{i,t}, D_{i,t+1}, \mathcal{G}_{i,t}, \mathcal{P}_{i,t}, \mathcal{A}_{i,t}, \mathcal{H}_{i,t}, \mathcal{V}_{i,t})$, where $g_{\mathrm{EC}}$ is a deterministic formatting map. The LLM may verbalize fields in $\mathcal{E}_{i,t}$ but may not introduce new forecasts, drivers, periods, or historical analogues. We then define three evidence interfaces. Let $\mathcal{R}_{i,t}$ denote the raw context containing the target series and retrieved windows, and let $\mathcal{S}_{i,t}$ denote a structured report plan. The interfaces are
\begin{equation}
\vspace{-1.5pt}
\mathcal{I}^{\mathrm{raw}}_{i,t}
=
\mathcal{R}_{i,t},
\quad
\mathcal{I}^{\mathrm{card}}_{i,t}
=
\mathcal{R}_{i,t}\oplus\mathcal{E}_{i,t},
\quad
\mathcal{I}^{\mathrm{str}}_{i,t}
=
\mathcal{R}_{i,t}\oplus\mathcal{E}_{i,t}\oplus\mathcal{S}_{i,t}.
\label{eq_evidence_interfaces_simple}
\vspace{-1.5pt}
\end{equation}
The operator $\oplus$ indicates that an additional structured object is supplied to the generator. The raw interface tests whether the LLM can select relevant evidence from unstructured context, the card interface adds the typed contract, and the structured-report interface further prescribes the reporting order. The comparison therefore measures the effect of {fixing evidence selection and document planning before generation}, not the effect of prompt length alone.

The structured report adds no factual information. It assigns existing evidence fields to report slots: formally, $\mathcal{S}_{i,t}=\{(s_g,A_g)\}_{g=1}^{G}$, where $s_g$ denotes a slot and $A_g\subseteq \mathcal{E}_{i,t}$ the fields permitted in that slot. The slots cover the summary judgment, forecast magnitude, baseline comparison, key drivers, temporal evidence, analogous cases, high-risk references, and caveats. Thus, the raw interface requires the LLM to decide both what to say and how to organize it, whereas the structured interface limits the model to verbalizing precomputed evidence within a fixed report grammar. This structure also clarifies the analyst's role. A free-form report forces reviewers to reconstruct the evidence path after generation, while a structured report exposes a finite set of typed claims that can be checked directly against the contract. The workflow reduces review burden without transferring the final risk judgment to the model.

\paragraph{Audit Protocol} \label{subsec_audit_protocol}
The audit layer verifies that a generated report remains within the evidence contract. Retrieval alone cannot ensure numerical, temporal, or role fidelity: a report may cite a relevant window but use the wrong unit, reverse the direction label, or present a high-risk reference as a reasoning-based analogue. The audit therefore evaluates hard factual validity before analyst review.

Let $G_{\phi}$ denote the LLM-based report generator. For an evidence interface $c\in\{\mathrm{raw},\mathrm{card},\mathrm{scaf}\}$, the generated report is $r^{c}_{i,t} = G_{\phi}(\mathcal{I}^{c}_{i,t}).$
Let $G_{\phi}$ denote the LLM-based report generator. For an evidence interface $c\in\{\mathrm{raw},\mathrm{card},\mathrm{scaf}\}$, the generated report is $r^{c}_{i,t} = G_{\phi}(\mathcal{I}^{c}_{i,t})$. The audit applies a finite set of checks to $r^{c}_{i,t}$. Let $\mathcal{Q}$ denote the set of direction, numerical-value, unit, feature, period, analogue, sample-support, and unsupported-inference checks. For each $q\in\mathcal{Q}$, the audit flag $\alpha_q(r^{c}_{i,t},\mathcal{E}_{i,t})$ is $1$ if $r^{c}_{i,t}$ violates $\mathcal{E}_{i,t}$ under check $q$, and $0$ otherwise. The audit output is $\boldsymbol{\alpha}(r^{c}_{i,t},\mathcal{E}_{i,t})=(\alpha_q(r^{c}_{i,t},\mathcal{E}_{i,t}))_{q\in\mathcal{Q}}$. This vector does not make an automated decision. So, it identifies the claims that require analyst verification.

Each check targets a specific failure mode. The direction check compares the stated increase, decrease, or flat classification with $e^{\mathrm{dir}}_{i,t}$. Numerical and unit checks verify forecasts, baselines, \%p changes, counts, and exposure quantities at the required precision. Feature and period checks ensure that every stated driver and date appears in $e^{\mathrm{drv}}_{i,t}$, $e^{\mathrm{time}}_{i,t}$, $e^{\mathrm{sim}}_{i,t}$, or $e^{\mathrm{risk}}_{i,t}$. The analogue check prevents reasoning-based analogues from being conflated with high-risk references, and the sample-support check retains required caveats from $e^{\mathrm{cav}}_{i,t}$. Finally, the unsupported-inference check flags causal, policy, institutional, or market claims that are not entailed by the contract.

The protocol is deliberately conservative: it verifies whether the draft preserves supplied evidence, while narrative quality and comparative model performance are evaluated separately. In deployment, the resulting flags define an explicit review surface on which an analyst can accept, revise, or reject the report.
\vspace{-2mm}

\section{Experiments} \label{sec_experiments}
\paragraph{Experiment Setting} \label{subsec_experiment_setting}

The empirical design separates two questions that are often conflated in financial monitoring. First, should the forecasting backbone be selected by average error or by its ability to surface right-tail accident observations that would otherwise be missed? Second, is explanation quality primarily determined by model scale, or does it also depend on how evidence is organized for the language model? Consistent with Section~\ref{method}, {the forecasting module identifies the warning case and produces the rationale matrices. Report generation is evaluated only after the forecast, direction label, retrieved precedents, and admissible evidence have been fixed.}

The forecasting sample is a monthly region-by-housing-type panel of guarantee completions, accident counts, collateral variables, housing-market indices, and macro-financial covariates. It contains 132 segment series over 124 months from September 2015 to December 2025. {We require a 25-month encoder with no more than one missing month and retain only windows with complete model inputs. This filtering yields 82 usable series, 4,580 windows, and 246 held-out test observations.} The target is the next-month guarantee accident rate. Appendix~\ref{app_forecasting_panel} consolidates sample construction, covariates, split, and tail-support definitions.Appendix~\ref{app_synthetic_material} documents {the synthetic reporting material} and Appendix~\ref{app_interface_contrast} specifies the matched interface estimands.

{For downstream LLM reporting, we separately construct synthetic aggregate explanation material so that source-level values are not exposed in prompts.} The forecasting backbone and model-selection logic {remain those estimated on the original panel}. The 24-case explanation stress test is drawn from {this LLM-facing material} and varies predicted risk, direction, exposure, forecast difficulty, and evidence availability. These cases are not used to estimate population forecasting accuracy. Instead, they are held fixed across models and interfaces to test whether reports preserve the same forecast, baseline, direction rule, model-derived drivers, CKA analogues, and high-risk references. Appendix~\ref{app_synthetic_material} {documents this synthetic reporting material and its separation from the confidential forecasting panel}.

The four questions, Q1--Q4, represent distinct tasks within the same monitoring report. Q1 and Q2 test recovery of the current target's driver and timing evidence, Q3 tests fidelity to a deterministic threshold rule rather than new forecasting ability, and Q4 tests whether the evidence can be assembled into a concise report without substituting analogous or high-risk observations for the current target.
\begin{itemize}[leftmargin=*,itemsep=0.15em,topsep=0.25em]
    \item \textbf{Q1, key driver.} Identify the main risk driver and support it with values or changes from the target segment.
    \item \textbf{Q2, important timing.} Identify the salient period in the target trajectory and compare it with the endpoint.
    \item \textbf{Q3, direction judgment.} {Compute the delta and direction label} from the recent baseline, forecast, and {$\pm 3.00$\%p threshold}, while preserving the label and numerical inputs.
    \item \textbf{Q4, structured report.} Produce a concise report that preserves the {direction label fixed in the prompt without recomputing or revising it}, numerical evidence, feature trajectory, analogous cases, high-risk references, and caveats.
\end{itemize}
We vary the evidence interface through three nested conditions. Raw Data supplies the target trajectory and retrieved windows. Evidence Card adds the typed contract, including direction inputs, feature trajectories, the salient period, analogous windows, high-risk references, and sample-support caveats. Structured Draft further adds a Q4 report grammar without introducing new facts. Raw Data and Evidence Card apply to Q1--Q4, whereas Structured Draft applies only to Q4. We therefore evaluate two matched supports: an all-question Raw-versus-Card contrast and a Q4-only three-interface contrast. This separation isolates evidence selection and document planning without treating unequal task distributions as directly comparable.

Eight LLMs answer Q1--Q4 under the admissible interfaces, producing 1,728 answer rows. The primary outcome is an answer-level report-quality score, complemented by deterministic rule flags and seven rubric axes: relevance, faithfulness, temporal reasoning, numerical grounding, consistency, domain validity, and hallucination control. Each axis is scored from 0 to 2, while the overall score is recorded separately from 0 to 10. Rule checks capture direction-label errors, key numerical errors, feature-flow omissions, unit-scale mistakes, placeholder leakage, and unsupported claims. Appendix~\ref{app_eval_points} summarizes the fixed-case design. Appendix~\ref{app_eval_rubric} defines the rubric and audit checks; Appendix~\ref{app_prompt_strategies} describes the three interfaces; Appendix~\ref{app_interface_contrast} defines the two matched estimands and Appendix~\ref{app_prompt} provides the fixed prompt governing direction, units, evidence use, and report structure.

\paragraph{Forecasting Backbone and Tail Monitoring} \label{subsec_forecast_results_exp}

The forecasting experiment varies both model family and training objective. LSTM, DLinear, LightGBM, and TFT provide the model comparison, while the TFT rows isolate conventional average-error objectives from the regret-sensitive objective. This distinction reflects the asymmetric monitoring problem: under-predicting a right-tail accident episode may delay review, whereas over-predicting a moderate segment primarily consumes analyst capacity \citep{christoffersen1997optimal,elliott2005estimation,gneiting2011making}.

Accordingly, Table~\ref{tab_forecast_perf_exp_} reports MAE, RMSE, tail recall, over-prediction frequency, and mean regret.
\begin{table}[ht!]
\centering
{\fontsize{8.5}{8.9}\selectfont
\setlength{\tabcolsep}{1.0pt}
\renewcommand{\arraystretch}{1.30}
\begin{tabular}{@{}llrrrrrr@{}}
\hline
Model & Obj. & MAE & RMSE & P70 & P90 & Over & Reg. $\times 100$ \\ \hline
LSTM & MAE & 0.027 & 0.091 & 0.743 & 0.080 & 0.659 & 0.0159 \\
DLinear & MAE & 0.039 & 0.092 & 0.851 & 0.440 & 0.642 & {0.0179} \\
LightGBM & MAE & 0.036 & 0.092 & {0.878} & 0.360 & {0.740} & {0.0176} \\
TFT & MAE & {0.029} & {0.092} & {0.419} & {0.160} & {0.317} & 0.0175 \\
TFT & RMSE & {0.034} & {0.091} & {0.554} & {0.400} & {0.500} & {0.0175} \\
\rowcolor[HTML]{DAE8FC}
\textbf{Ours} & \textbf{Regret} & \textbf{{0.041}} & \textbf{{0.097}} & \textbf{{0.703}} & \textbf{{0.560}} & \textbf{{0.598}} & \textbf{{0.0179}} \\
\hline
\end{tabular}}
\caption{Out-of-sample performance for next-month guarantee accident rates on 246 test observations. P70 and P90 denote recall within the realized upper 30\% and upper 10\% tails, Over is the share of predictions exceeding realizations, and Reg. $\times 100$ is scaled asymmetric regret.}
\label{tab_forecast_perf_exp_}
\vspace{-15pt}
\end{table}
Table~\ref{tab_forecast_perf_exp_} reveals a clear divergence between average accuracy and tail monitoring. LSTM-MAE achieves the lowest MAE, 0.027, yet its P90 recall is only 0.080, meaning that it surfaces fewer than one in ten realized upper-tail observations. Thus, the strongest conventional accuracy benchmark remains weak precisely where missed warnings are most costly. This result reflects the zero-heavy test distribution: 54.1\% of observations are zero, the median is zero, and the P90 cutoff is 0.069.

The model-family baselines further characterize this tradeoff. DLinear and LightGBM increase P70 recall to 0.851 and {0.878}, respectively, but neither directly yields the variable-time rationale matrix required by the retrieval and evidence-card layers. Their P90 recalls, 0.440 and 0.360, also remain below the selected backbone. Within TFT, the MAE-trained model records MAE of {0.029} and P90 recall of {0.160}, while the RMSE-trained model records RMSE of {0.091} and P90 recall of {0.400}. These comparisons show that an interpretable architecture is not sufficient by itself. Hence, the objective must also shift the operating point toward missed-tail reduction.

The Ours row reports the operating point used for evidence generation. TFT-Regret records MAE of {0.041}, RMSE of {0.097}, P70 recall of {0.703}, and P90 recall of {0.560}. On the common 246-observation test set, it surfaces {14 of the 25} realized P90 cases, compared with {2 of 25} for LSTM-MAE, {11 of 25} for DLinear-MAE, and {9 of 25} for LightGBM-MAE. This gain costs {1.37\%p} in MAE relative to the lowest-MAE model. Because the next stage is analyst review rather than automatic intervention, the tradeoff is operationally coherent: it expands the high-risk review set while retaining the TFT rationale matrices needed for CKA retrieval and evidence construction. Appendix~\ref{app_tft_objective} reports the operating frontier as the under-prediction penalty increases from $\lambda=3$ to $\lambda=10$. Higher penalties improve tail recall but also increase average error and over-prediction. 

\paragraph{Evidence Interfaces and Explanation Quality} \label{subsec_llm_results_exp}

The second empirical question concerns whether report fidelity is determined by model capability alone or also by the organization of admissible evidence. Because Structured Draft is defined only for Q4, placing Raw Data, Evidence Card, and Structured Draft in a single model table would compare different task supports. We therefore report two matched estimands. The first compares Raw Data with Evidence Card over the same 24 cases, four questions, and eight LLMs. The second restricts all three interfaces to Q4, where the structured report plan is admissible.
\begin{table}[ht!]
\centering
\resizebox{\columnwidth}{!}{%
{\fontsize{8.8}{9.8}\selectfont
\setlength{\tabcolsep}{3.0pt}
\renewcommand{\arraystretch}{1.08}
\begin{tabular}{@{}lrrrrrrrr@{}}
\hline
Interface & \cellcolor[HTML]{DAE8FC}GPT-4o & \makecell[c]{Upstage\\Solar} & \makecell[c]{Mistral\\24B} & \makecell[c]{Llama\\70B} & \makecell[c]{Hyper\\CLOVA X} & \makecell[c]{Local\\Base 8B} & \makecell[c]{Local Time\\-MQA 8B} & \makecell[c]{EXAONE\\Local} \\ \hline
Raw & \cellcolor[HTML]{DAE8FC}\textbf{{8.56}} & {8.19} & {7.97} & {7.75} & {5.79} & {3.07} & {3.10} & {2.15} \\
Evid. & \cellcolor[HTML]{DAE8FC}\textbf{{8.77}} & {8.39} & {8.20} & {8.25} & {7.60} & {4.38} & {4.01} & {4.09} \\
\hline
\end{tabular}%
}}
\caption{Q1--Q4 matched mean report-quality scores by LLM. Raw Data and Evidence Card are evaluated on the same 24 cases and four questions, so each model-interface cell averages 96 answers.}
\label{tab_model_strategy_exp_}
\vspace{-22pt}
\end{table}
Table~\ref{tab_model_strategy_exp_} shows that evidence organization contributes beyond model choice. Evidence Card improves every model relative to Raw Data: GPT-4o rises from {8.56} to {8.77}, Upstage Solar from {8.19} to {8.39}, Mistral 24B from {7.97} to {8.20}, Llama 70B from {7.75} to {8.25}, and HyperCLOVA X from {5.79} to {7.60}. The local models also improve, with Local Base 8B increasing from {3.07} to {4.38}, Local Time-MQA 8B from {3.10} to {4.01}, and EXAONE Local from {2.15} to {4.09}. Averaged over the same Q1--Q4 support, the score increases from {5.82} to {6.71}.

For the complete reporting task, Table~\ref{tab_prompt_rule_exp_} restricts the comparison to Q4 and reports deterministic pass rates available on the same case-model support for all three interfaces.
\begin{table}[ht!]
\centering
{\fontsize{8.8}{8.8}\selectfont
\setlength{\tabcolsep}{2.4pt}
\renewcommand{\arraystretch}{1.08}
\begin{tabular}{@{}lrrrr@{}}
\hline
Interface & Top feat. & Timing & Dir. values & High-risk \\ \hline
Raw Data & {90.6} & {21.9} & {70.8} & {77.1} \\
Evidence Card & {98.4} & {62.0} & {87.5} & {86.5} \\
Structured Draft & {100.0} & {90.6} & {100.0} & {91.7} \\
\hline
\end{tabular}}
\caption{Q4-only matched diagnostic pass rates by evidence interface (\%).}
\label{tab_prompt_rule_exp_}
\vspace{-12pt}
\end{table}

The Q4-only results identify the mechanism behind the interface gains. Evidence Card raises top-feature preservation from {90.6\%} to {98.4\%}, salient-period mention from {21.9\%} to {62.0\%}, exact direction-value preservation from {70.8\%} to {87.5\%}, and correct high-risk-reference handling from {77.1\%} to {86.5\%}. Structured Draft further increases the corresponding rates to {100.0\%}, {90.6\%}, {100.0\%}, and {91.7\%}. Under this Q4-only condition, final scores are {8.71} for GPT-4o, {8.54} for Llama 70B, {8.04} for Upstage Solar, {7.96} for Mistral 24B, {7.67} for HyperCLOVA X, {7.42} for Local Base 8B, {7.33} for Local Time-MQA 8B, and {6.83} for EXAONE Local. These values are reported as Q4 endpoints rather than pooled with Q1--Q3.

Appendix~\ref{app_interface_contrast} defines the paired supports, Appendix~\ref{app_exp_details} documents the experimental design, and Appendix~\ref{app_prompt_strategies} specifies the nested interfaces. Raw Data requires the LLM to locate relevant values and determine report order; {Evidence Card fixes evidence selection before generation; and Structured Draft additionally fixes Q4 document planning.} The reductions in rule violations therefore reflect factual control rather than a change in task composition or a cosmetic improvement in prose.

\begin{figure*}[t]
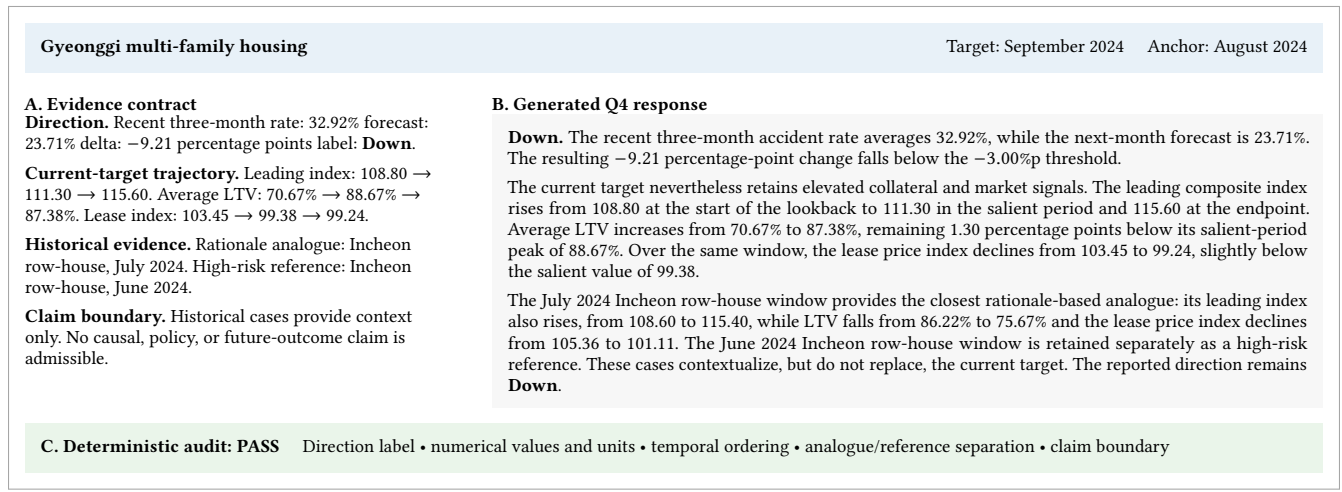

\centering
\setlength{\fboxsep}{6pt}
\fcolorbox{black!35}{white}{%
\begin{minipage}{0.965\textwidth}
\footnotesize
\noindent\colorbox[HTML]{E8F1F8}{%
\parbox{\dimexpr\linewidth-2\fboxsep\relax}{\textbf{Gyeonggi multi-family housing} \hfill Target: September 2024 \quad Anchor: August 2024}}
\vspace{5pt}

\begin{minipage}[t]{0.32\linewidth}
\raggedright
\textbf{A. Evidence contract}\\[-1pt]
\textbf{Direction.} Recent three-month rate: 32.92\% forecast: 23.71\% delta: $-9.21$ percentage points label: \textbf{Down}.\\[3pt]
\textbf{Current-target trajectory.} Leading index: 108.80 $\rightarrow$ 111.30 $\rightarrow$ 115.60. Average LTV: 70.67\% $\rightarrow$ 88.67\% $\rightarrow$ 87.38\%. Lease index: 103.45 $\rightarrow$ 99.38 $\rightarrow$ 99.24.\\[3pt]
\textbf{Historical evidence.} Rationale analogue: Incheon row-house, July 2024. High-risk reference: Incheon row-house, June 2024.\\[3pt]
\textbf{Claim boundary.} Historical cases provide context only. No causal, policy, or future-outcome claim is admissible.
\end{minipage}\hfill
\begin{minipage}[t]{0.64\linewidth}
\raggedright
\textbf{B. Generated Q4 response}\\[-1pt]
\colorbox[HTML]{F7F7F7}{%
\parbox{\dimexpr\linewidth-2\fboxsep\relax}{%
\textbf{Down.} The recent three-month accident rate averages 32.92\%, while the next-month forecast is 23.71\%. The resulting $-9.21$ percentage-point change falls below the $-3.00$\%p threshold.

\medskip
The current target nevertheless retains elevated collateral and market signals. The leading composite index rises from 108.80 at the start of the lookback to 111.30 in the salient period and 115.60 at the endpoint. Average LTV increases from 70.67\% to 87.38\%, remaining 1.30 percentage points below its salient-period peak of 88.67\%. Over the same window, the lease price index declines from 103.45 to 99.24, slightly below the salient value of 99.38.

\medskip
The July 2024 Incheon row-house window provides the closest rationale-based analogue: its leading index also rises, from 108.60 to 115.40, while LTV falls from 86.22\% to 75.67\% and the lease price index declines from 105.36 to 101.11. The June 2024 Incheon row-house window is retained separately as a high-risk reference. These cases contextualize, but do not replace, the current target. The reported direction remains \textbf{Down}.
}}
\end{minipage}

\vspace{5pt}
\noindent\colorbox[HTML]{EAF5EA}{%
\parbox{\dimexpr\linewidth-2\fboxsep\relax}{\textbf{C. Deterministic audit: PASS} \quad Direction label \textbullet\ numerical values and units \textbullet\ temporal ordering \textbullet\ analogue/reference separation \textbullet\ claim boundary}}
\end{minipage}}
\caption{Audited Q4 report showcase for Case 24. Panel A exposes the typed evidence available to the generator. Panel B reproduces the analyst-facing response. Panel C summarizes {diagnostic checks}. The analogue and high-risk reference remain distinct, and neither replaces target-specific evidence.}
\Description{Full-width qualitative example with an evidence contract, a generated housing-guarantee risk report, and a {diagnostic audit strip} showing that direction, numerical, temporal, evidence-role, and claim-boundary checks pass.}
\label{fig_q4_showcase}
\vspace{-5pt}
\end{figure*}

\paragraph{Audited Report Showcase} \label{subsec_q4_showcase}
Figure~\ref{fig_q4_showcase} instantiates one held-out Q4 case. Panel A fixes the admissible facts, Panel B shows the Structured Draft response, and Panel C records the {diagnostic audit}. The report preserves the supplied direction, values, units, temporal order, and the distinct roles of analogous and high-risk windows. It also illustrates that \textit{Down} is relative to the recent baseline: elevated LTV and a weakening lease-price index remain visible for review. In contrast, low-scoring outputs reversed the direction, omitted the high-risk reference, or introduced unsupported LTV values. Appendix~\ref{app_prompt} provides the fixed preventive instructions.

\paragraph{Evaluation Decomposition} \label{subsec_eval_diagnostics_exp}

\begin{figure}[ht!]
\centering
\includegraphics[width=0.8\linewidth]{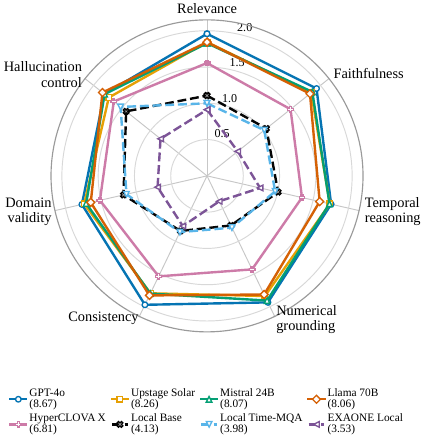}
\caption{Descriptive seven-axis report-quality profiles pooled over each model's evaluated answers. Matched interface effects are reported separately in Tables~\ref{tab_model_strategy_exp_} and~\ref{tab_prompt_rule_exp_}.}
\Description{Seven-axis chart comparing evaluation dimensions across models.}
\label{fig_eval_all_exp}
\vspace{-10pt}
\end{figure}

Figure~\ref{fig_eval_all_exp} provides a descriptive profile pooled over each model's evaluated answers. It is not used to estimate an interface effect. Those effects are reported on matched supports in Tables~\ref{tab_model_strategy_exp_} and~\ref{tab_prompt_rule_exp_}. The strongest models are not merely more fluent: they also perform better on numerical grounding, faithfulness, and hallucination control, the dimensions most directly governed by the evidence contract. This pattern supports the audit protocol in Section~\ref{subsec_audit_protocol}, under which operational reports must preserve direction thresholds, accident rates, lookback periods, evidence roles, and caveats.

The pooled profiles of the local models reveal weaknesses in relevance, numerical grounding, and consistency. These dimensions correspond directly to factual review cost. Analysts can revise style locally, whereas a reversed direction label or fabricated accident count requires returning to the underlying evidence. The decomposition therefore clarifies the review labor that structured evidence is intended to reduce.

\begin{figure}[ht!]
\centering
\includegraphics[width=0.78\linewidth]{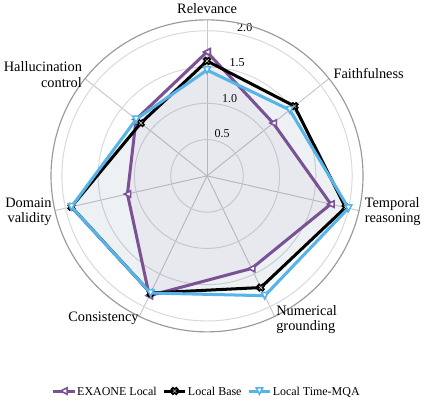}
\caption{Seven-axis Q4 Structured Draft profiles for local and private LLM candidates. Structured evidence most clearly improves temporal reasoning and consistency, while faithfulness and hallucination control remain the principal weaknesses.}
\Description{Seven-axis chart comparing Q4 structured-draft scores for local and private model candidates.}
\label{fig_eval_local_exp}
\vspace{-22pt}
\end{figure}

\paragraph{Local and Private Model Diagnostics}
The Q4-only results sharpen the deployment interpretation for local and private candidates. Figure~\ref{fig_eval_local_exp} shows that Local Base 8B and Local Time-MQA 8B approach or exceed 1.6 in temporal reasoning and consistency under Structured Draft, while EXAONE Local also improves enough to support bounded drafting. Faithfulness and hallucination control remain the principal residual weaknesses, consistent with the Q4 rule diagnostics in Table~\ref{tab_prompt_rule_exp_}.

The magnitude of the matched Q4 Raw-to-Draft change is operationally important. Local Base 8B increases from {4.25} to {7.42}, Local Time-MQA 8B from {4.29} to {7.33}, and EXAONE Local from {2.38} to {6.83}. These results do not justify autonomous local reporting. They indicate that institutions restricting external API use may consider a controlled workflow in which local models draft from pre-validated evidence and {deterministic checks and judge-assisted diagnostics expose residual numerical or grounding errors}.

\paragraph{Real-world Deployment}
We further examine whether the report functions as an operational review aid beyond the synthetic LLM stress test. A representative evidence-grounded report was distributed to 51 analysts and domain practitioners working in guarantee review, accident management, strategy, fund and systems, and data and AI roles. Respondents rated seven task-specific statements and overall usefulness on a five-point scale. Because the exercise {did not randomly assign practitioners to treatment and control conditions or embed the report in live operational decisions}, we interpret it as direct end-user validation rather than a causal estimate of productivity or decision quality.
\begin{figure}[!t]
\centering
\includegraphics[width=0.98\linewidth]{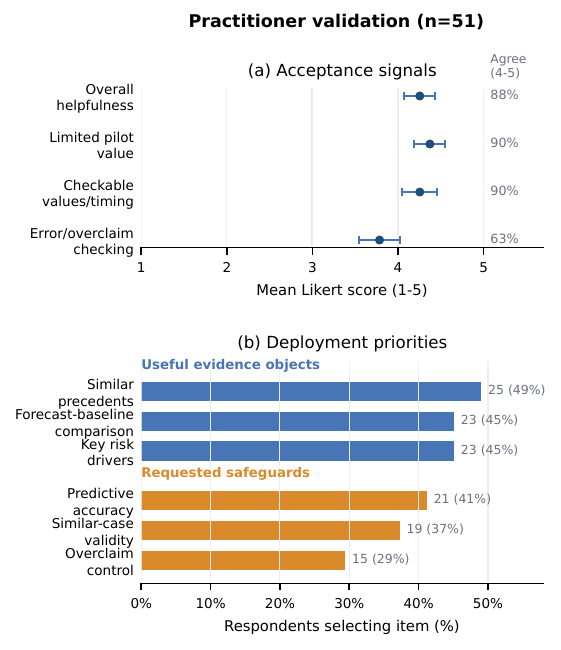}
\caption{Post-deployment practitioner evaluation of the evidence-grounded reporting interface. Panel (a) shows selected five-point acceptance measures with 95\% normal-approximation intervals and agreement shares. Panel (b) shows useful evidence objects in blue and requested safeguards in orange. Appendix~\ref{app_realworld} provides the full distributions, respondent profile, and multi-select counts.}
\Description{Compact two-panel practitioner survey figure showing acceptance signals and deployment priorities.}
\label{fig_realworld_main}
\vspace{-7mm}
\end{figure}
Figure~\ref{fig_realworld_main} shows three deployment-relevant findings. Overall helpfulness averages {4.25} out of 5, with {88.2\%} of respondents selecting 4 or 5. Support for a pilot is strongest, with {90.2\%} agreeing that the report merits testing in a constrained pre-deployment workflow. Checkability also receives {90.2\%} agreement and a mean of {4.25}. By contrast, the ability to detect explanation errors or overstatements, while still positive, is the lowest-rated item at {3.78} and {62.7\%} agreement. Practitioners therefore value visible evidence but continue to require explicit verification rather than autonomous narrative generation.

The component responses identify the evidence and safeguards that should be retained in deployment. The most useful objects are similar historical precedents ({49.0\%}), the forecast-baseline comparison ({45.1\%}), key risk drivers ({45.1\%}), and numerical feature changes ({43.1\%}), all of which are fixed by the contract before generation. The most requested safeguards are improved predictive accuracy ({41.2\%}), validation of similar-case relevance ({37.3\%}), overstatement control ({29.4\%}), data-source labels ({27.5\%}), and automatic numerical-error checks ({25.5\%}). Open-text responses additionally request tables, graphs, and clearer interpretation criteria.

\section{Lessons Learned} \label{sec_lessons}
Our central lesson from the analyst-facing deployment is that operational value came from fitting established methods to the institution's review capacity, data-governance rules, and accountability structure rather than from any single novel component.
\vspace{-2mm}

\paragraph{Optimize for the review decision.}
Housing-guarantee monitoring is a queue-allocation problem as well as a forecasting problem. The lowest-MAE model detected only 2 of 25 P90 observations, whereas TFT-Regret at $\lambda=3$ detected 14, at the cost of higher MAE and over-prediction. Larger penalties recovered more tail cases but expanded the review queue. We therefore selected an operating point that balanced tail visibility with analyst capacity and kept the forecast advisory. In sparse, zero-heavy panels, model selection should report both central accuracy and the review workload induced by tail-sensitive detection.
\vspace{-2.1mm}

\paragraph{Move evidence selection outside the LLM.}
The Raw Data condition exposed factual rather than stylistic failures: responses reversed direction labels, confused target and historical windows, omitted salient periods, and altered values or units. We therefore separated rationale analogues from outcome-based high-risk references, encoded admissible facts in the Evidence Card, and used Structured Draft only to fix Q4 document order. The Evidence Card improved every evaluated model on the matched Q1--Q4 support, while Structured Draft further improved Q4 diagnostics and made smaller local models viable bounded drafters. The practical lesson is to use the LLM to verbalize pre-validated evidence, not to reconstruct decision logic from raw tables.
\vspace{-2.1mm}

\paragraph{Confidentiality defines the deployment boundary.}
The forecasting layer remained inside the secure environment, while the analyst-facing reporting layer exposed only the bounded evidence required for review. {We used internally consistent synthetic aggregates for reproducible LLM evaluation, while the forecasting backbone and model-selection logic remained estimated on the original panel inside the secure environment.} This separation enabled analyst-facing deployment without transmitting source-level records or sensitive regional aggregates to external models. In policy-sensitive settings, data minimization, local-model compatibility, and disclosure control are deployment requirements rather than post hoc safeguards.
\vspace{-2.1mm}

\paragraph{Deploy the workflow, not the model.}
Fluent reports still contained unsupported claims, making {deterministic checks, judge-assisted diagnostics,} and analyst review necessary parts of the deployed system. We deployed the evidence-grounded reporting interface to 51 analysts and related practitioners in a controlled operational evaluation. Their feedback indicated that checkable evidence, historical precedents, and forecast--baseline comparisons supported review, while error detection and precedent validity remained the principal safeguards. The deployment was therefore intentionally bounded to analyst decision support: the complete forecast--retrieval--contract--generation--audit workflow assisted review, while final judgments and policy actions remained under existing institutional controls.

\section{Conclusion}

This paper presents a controlled analyst-facing deployment of an evidence-constrained workflow for converting housing-guarantee forecasts into auditable reports under confidentiality and review constraints. A regret-sensitive TFT improves upper-tail visibility, while Evidence Cards and Q4 Structured Drafts reduce factual failures by fixing admissible evidence before generation. In an operational evaluation with 51 analysts and related practitioners, the reporting interface was assessed as useful for review and valued for its checkable evidence, while the feedback reinforced the need for explicit audits and precedent validation. The contribution is an analyst-centered architecture that combines decision-aligned forecasting, rationale-based retrieval, typed evidence, {deterministic checks, judge-assisted diagnostics,} and human approval, with final judgments and policy actions retained under institutional control.

\section*{Disclaimer}
The views and opinions expressed in this paper are solely those of the authors and do not represent the official position, policies, or views of the Korea Housing \& Urban Guarantee Corporation (HUG). Any interpretations, conclusions, recommendations, errors, or omissions are the sole responsibility of the authors. 


\vspace{2\baselineskip}
\bibliographystyle{ACM-Reference-Format}
\bibliography{sample-base}

\clearpage
\appendix
\onecolumn

\section{Experimental Design and Reproducibility} \label{app_exp_details}

\subsection{Forecasting panel, covariates, and split} \label{app_forecasting_panel}

The panel contains 132 region-by-housing-type series over 124 months. For segment $i$ and anchor $t$, a 25-month encoder predicts the next-month accident rate $y_{i,t+1}=\frac{\text{accident count}_{i,t+1}}{\text{completed guarantee count}_{i,t+1}}$. {A complete encoder, a positive target denominator, at most one missing encoder month, and complete model inputs after construction are required.} These criteria leave 82 series and 4,580 windows: 4,088 training, 246 validation, and 246 test. Reports use the encoder endpoint as the anchor, so a September 2024 target is reported from August 2024. The encoder uses 13 lagged inputs. Guarantee activity is represented by completed guarantee count, average guarantee amount, and average maturity. Collateral conditions by average housing value, average LTV, and the share of guarantees with LTV above 80\%, housing-market conditions by lease and sale price indices and macro-financial conditions by the policy rate, CPI, unemployment rate, leading composite index, and promissory-note default rate. These groups describe operational roles rather than causal structure, and selected variables are treated only as model-rationale evidence. P70 and P90 recall apply the realized 70th- and 90th-percentile cutoffs to both forecasts and outcomes. Their test supports are 74 and 25 observations. Over is the share for which $\widehat y_{i,t+1}>y_{i,t+1}$, and regret is evaluated by Eq.~\ref{eq_regret_loss} and scaled by 100.

\subsection{Confidential reporting material and robustness} \label{app_synthetic_material}

The forecasting backbone is estimated on confidential records, whereas the LLM-facing prompts use synthetic aggregates because regional summaries may remain sensitive. The synthetic material preserves the schema, units, 25-month encoder geometry, empirical support, and temporal variation of the original panel. Rates, recent baselines, deltas, and labels under the $\pm3.00$ percentage-point rule are recomputed, and target windows remain distinct from rationale-based analogues and outcome-based high-risk references. The pool contains 374 windows, from which 24 validation/test cases are purposively selected to span risk, direction, exposure, forecast difficulty, and evidence availability.

{This construction separates the confidential forecasting experiment from the LLM-facing reporting experiment. The original panel determines the forecasting backbone and model-selection logic, while the synthetic material determines only the prompt-visible evidence for report generation. The comparison tests whether LLMs preserve fixed forecasts, direction labels, numerical values, retrieved precedents, and caveats without disclosing production aggregates.}

\subsection{Matched interface contrasts} \label{app_interface_contrast}

Q1--Q3 use Raw Data and Evidence Card, whereas Q4 additionally uses Structured Draft. The design therefore contains $24\times(3\times2+1\times3)=216$ conditions, or 1,728 answers across eight LLMs. For score $Y_{s,q,m,c}$, the paired non-raw contrast is $\widehat{\Delta}_{c} = \frac{1}{|\Omega_c|} \sum_{(s,q,m)\in\Omega_c} \left(Y_{s,q,m,c} -Y_{s,q,m,\mathrm{raw}}\right)$ where $\Omega_c$ fixes the case, question, model, forecast, and retrieval set. For Evidence Card, $\Omega_{\mathrm{card}}$ contains Q1--Q4; for Structured Draft, $\Omega_{\mathrm{draft}}$ contains Q4 only. The two main-text tables follow these respective supports, and the contrasts are descriptive rather than causal.

\section{TFT Objective and Operating-Point Selection} \label{app_tft_objective}

The under-prediction penalty $\lambda$ raises tail recall as it increases, but also raises average error and over-prediction. Table~\ref{tab_app_tft_objective} reports this operating frontier.

\begin{table}[H]
\centering
\caption{TFT objective comparison on the common 246-observation test set. P70 and P90 are realized-tail recall, Over is the share of predictions exceeding realizations, and P90 hits reports detections among the 25 P90 cases.}
\label{tab_app_tft_objective}
\scriptsize
\setlength{\tabcolsep}{4.3pt}
\renewcommand{\arraystretch}{1.08}
\begin{tabular}{llrrrrrrr}
\toprule
Model & Objective & MAE & RMSE & P70 & P90 & P90 hits & Over & Regret $\times100$ \\
\midrule
TFT & MAE & {0.029} & {0.092} & {0.419} & {0.160} & 4 & {0.317} & 0.0175 \\
TFT & RMSE & {0.034} & {0.091} & {0.554} & {0.400} & 10 & {0.500} & {0.0175} \\
TFT & Regret $\lambda=3$ & \textbf{{0.041}} & \textbf{{0.097}} & \textbf{{0.703}} & \textbf{{0.560}} & \textbf{14} & \textbf{{0.598}} & \textbf{{0.0179}} \\
TFT & Regret $\lambda=5$ & {0.068} & {0.134} & {0.892} & {0.600} & 15 & {0.821} & {0.0242} \\
TFT & Regret $\lambda=10$ & {0.103} & {0.177} & {0.986} & {0.880} & 22 & {0.915} & {0.0338} \\
\bottomrule
\end{tabular}
\end{table}

We select the first regret checkpoint because it surfaces 14 P90 cases with materially less review load than the $\lambda=5$ and $\lambda=10$ settings.

\section{LLM Stress-Test Design} \label{app_eval_points}

The 24 fixed cases are drawn from the validation/test portion of the synthetic pool and held constant across models and admissible interfaces. They comprise {8 low-, 8 medium-, and 8 high-risk cases; 8 small-, 8 medium-, and 8 large-exposure cases; 15 Down, 6 Flat, and 3 Up cases; and 8 low-, 8 medium-, and 8 high-error cases}. Dominant evidence is distributed across the promissory-note default rate in 4 cases, the leading composite index in 10, and the lease price index in 10. Predicted accident rates range from 2.14\% to 30.45\%.

Each case fixes the target and anchor month, forecast, recent baseline, threshold inputs, current feature trajectory, rationale-based analogues, outcome-based high-risk references, and sample-support caveats. The target month is one month after the report anchor. This fixed design tests evidence preservation rather than population forecast accuracy, and sensitive pool cutoffs are withheld.

\section{Evidence Interfaces and Generation Prompt} \label{app_prompt_strategies}

\paragraph{Nested interfaces.}
Raw Data supplies the target 25-month trajectory, direction inputs, forecast values, and retrieved historical windows with minimal organization. Evidence Card adds typed direction fields, the recent baseline, forecast, delta, threshold, selected feature trajectories, salient period, analogous windows, high-risk references, counts, and sample-support caveats. Structured Draft adds no facts. It assigns those fields to a Q4 report grammar that fixes document order, opening and closing direction statements, feature-flow discussion, separation of analogue and high-risk evidence, and caveat placement. Raw Data and Evidence Card apply to Q1--Q4, while Structured Draft applies only to Q4. In every matched comparison, the target, forecast, retrieval set, question, and model identity remain fixed. Analogues are selected by rationale similarity and high-risk references by realized outcomes. Neither may replace target-specific evidence.

\subsection{Fixed generation prompt} \label{app_prompt}

Figure~\ref{fig_app_prompt} reproduces the common instruction block. Only case-specific evidence slots vary across calls. Q3 tests rule execution, whereas Q4 tests exact preservation of validated values in a longer report. Only final answers are scored.

\begin{figure}[H]
\centering
\setlength{\fboxsep}{6pt}
\fbox{%
\begin{minipage}{0.95\linewidth}
\small
\textbf{Prompt. Evidence-grounded guarantee accident risk report}

\textbf{Common direction rule.}
The next-month guarantee accident risk direction is determined by the difference between the recent three-month average accident rate and the next-month predicted accident rate.
Let \texttt{recent3\_rate} be the recent three-month average accident rate. Let \texttt{prediction\_rate} be the next-month predicted accident rate. Let \texttt{delta = prediction\_rate - recent3\_rate}. The threshold is $\pm 3.00$ percentage points. If \texttt{delta} is at least $+3.00$ percentage points, the direction is \texttt{Up}. If \texttt{delta} is at most $-3.00$ percentage points, the direction is \texttt{Down}. If \texttt{delta} lies strictly between $-3.00$ and $+3.00$ percentage points, the direction is \texttt{Flat}. Do not reverse the inequality for negative values. The direction label is a comparison with recent observed levels and not an absolute risk ranking.

\textbf{Target answer rule.}
Write the answer for the current target segment, housing type, and anchor month only. For Q3, compute \texttt{delta} from the supplied \texttt{recent3\_rate} and \texttt{prediction\_rate}, then apply the $\pm 3.00$\%p threshold. For Q4, use the supplied \texttt{recent3\_rate}, \texttt{prediction\_rate}, \texttt{delta}, and \texttt{direction\_label} exactly as given. Do not replace these values with accident rates from the raw time series, similar cases, or high-risk reference windows. Similar cases and high-risk windows are comparative evidence only, and their region, month, and accident rate must not be substituted for the current target.

\textbf{Domain background.}
The guarantee accident rate is the ratio of accident count to completed guarantee count at the monthly segment level. All reported quantities are monthly aggregate averages, counts, or ratios rather than individual contract values. A higher LTV indicates a larger guarantee burden relative to collateral value. Lease and sale price indices are auxiliary housing-market indicators. Macro variables are auxiliary market-condition signals and must not be stated as direct causes. When accident count and guarantee count are used to express a rate, convert ratios to\% correctly. For example, $51/1{,}781=0.0286=2.86\%$, not $0.0286\%$.

\textbf{Evidence interface.}
Use the raw target time series and the retrieved historical windows only as supplied. Under the Evidence Card condition, first use the provided key feature, important lookback period, direction inputs, similar cases, high-risk reference window, accident count, guarantee count, and caveats. Under the Structured Draft condition, follow the given report order but do not introduce new facts. Do not use internal terms such as TFT, CKA, attention, feature weight, prompt cache, or placeholder tokens in the final answer.

\textbf{Question-specific instructions.}
For Q1, state the main driver first and support it with at least two numerical values or changes from the current target. For Q2, identify the important period for the current target and compare its value with the terminal period. For Q3, give the direction label after computing \texttt{delta = prediction\_rate - recent3\_rate}, and briefly cite the two rates, the computed delta, and the $\pm 3.00$\%p threshold. For Q4, write an expert report in no more than five paragraphs. The first and last sentences must use the supplied direction label. The report must connect the direction inputs, key feature trajectory, similar historical cases, high-risk reference window, and caveats.

\textbf{Q4 pre-answer checklist.}
Before writing Q4, verify the recent three-month average, predicted rate, supplied delta, and threshold. Do not recalculate or change the supplied direction label. Present the main feature with start, salient, and terminal values when available. Separate similar cases from high-risk references. If a historical case has a small guarantee count, state that the accident rate is sensitive to sample support. Do not invent future realized accident rates, policy events, accident counts, or unprovided regional facts. Return only the final expert report.
\end{minipage}}
\caption{Fixed prompt for evidence-grounded report generation. Target-specific fields are populated before each call, while the instruction logic remains constant across models and admissible interfaces.}
\Description{Boxed text of the fixed prompt used for report generation.}
\label{fig_app_prompt}
\end{figure}

\section{Evaluation Rubric and Audit Checks} \label{app_eval_rubric}

Answers are scored against the case evidence rather than model identity. Seven axes receive 0--2 points. Relevance measures whether the response directly addresses the requested task. Faithfulness assesses whether it uses only the supplied evidence without altering its meaning. Temporal reasoning evaluates whether the start, salient period, terminal point, anchor, and target remain distinct. Numerical grounding checks the accuracy of rates, deltas, feature values, and counts. Consistency examines whether the direction label remains aligned throughout the report. Domain validity assesses whether accident rates, LTV, price indices, and sample support are interpreted without causal overstatement. Hallucination control evaluates whether the response avoids placeholders, future outcomes, and unsupported external facts. A score of 0 indicates absence or a material error, 1 indicates partial or ambiguous fulfillment, and 2 indicates complete and supported fulfillment. The separate 0--10 overall score is holistic and is not a rescaling of the seven axes.

{The diagnostic audit combines deterministic rule checks with judge-assisted unsupported-claim diagnostics as summarized in Table~\ref{tab_app_rule_flags_compact}.} Each flag identifies a factual failure that requires analyst review and remains visible rather than being averaged away by fluent prose or a high holistic score.

\begin{table}[H]
\centering
\caption{{Diagnostic audit checks and their applicable tasks.}}
\label{tab_app_rule_flags_compact}
\scriptsize
\setlength{\tabcolsep}{4pt}
\renewcommand{\arraystretch}{1.08}
\begin{tabular}{p{0.20\linewidth}p{0.57\linewidth}p{0.14\linewidth}}
\toprule
Check & Failure condition & Tasks \\
\midrule
Direction label & Up, Flat, or Down conflicts with the supplied label or the Q3 threshold calculation. & Q3--Q4 \\
Key numerical value & A baseline, forecast, delta, feature value, count, or reference rate is missing or inconsistent. & Q1--Q4 \\
Feature flow & Start, salient, or terminal evidence is omitted or assigned to the wrong series or period. & Q1, Q2, Q4 \\
Unit and scale & Ratios, percentages, percentage points, counts, or index levels are converted or labeled incorrectly. & Q1--Q4 \\
Placeholder leakage & Internal instructions, model terms, or unresolved drafting tokens remain in the answer. & Q4 \\
Unsupported claim & A causal, policy, market, institutional, or future-outcome statement is not entailed by the contract. & Q1--Q4 \\
\bottomrule
\end{tabular}
\end{table}

\section{Real-World Practitioner Evaluation} \label{app_realworld}

\begin{figure}[H]
\centering
\includegraphics[width=0.96\linewidth]{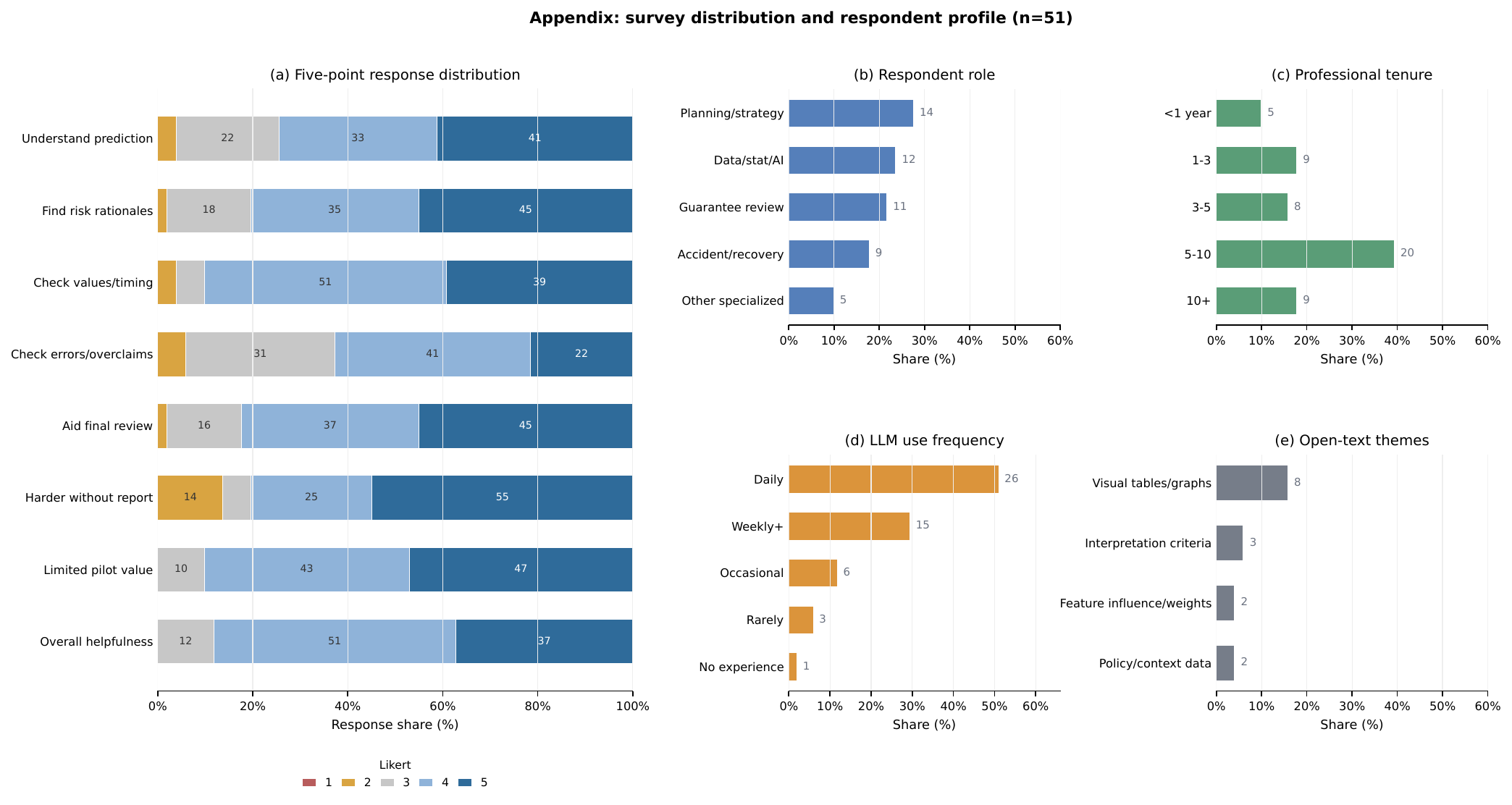}
\caption{Practitioner response distributions, respondent profiles, and open-text themes.}
\Description{Appendix survey figure with Likert distributions, respondent roles, tenure, LLM use, and open-text themes.}
\label{fig_realworld_appendix}
\end{figure}

Fifty-one practitioners rated seven task-specific statements and overall helpfulness on five-point scales. The pool includes 14 planning, 12 data/AI, 11 screening, 9 accident/debt-management, and 5 specialized-function respondents; 37 have at least three years of experience and 41 use LLMs weekly. Figure~\ref{fig_realworld_appendix} reports the complete response distributions, respondent profile, and coded open-text themes. The 95\% intervals use $\widehat p\pm1.96\sqrt{\widehat p(1-\widehat p)/51}$ and describe this non-random pool rather than a population estimate.

Across the eight rated statements, means range from 3.78 to 4.37 and agreement from 62.7\% to 90.2\%. Overall helpfulness averages 4.25, with 88.2\% agreement, pilot support averages 4.37, with 90.2\% agreement, and checkable presentation averages 4.25, also with 90.2\% agreement. Error or overstatement detection is the weakest item, at 3.78 and 62.7\% agreement, reinforcing the need for explicit audits.

The most frequently selected evidence objects are similar historical precedents (25 respondents), the forecast-baseline comparison (23), key risk drivers (23), and numerical feature changes (22). The leading requested safeguards are improved predictive accuracy (21), validation of similar-case relevance (19), overstatement control (15), data-source labels (14), and automatic numerical-error checks (13). Multi-select shares need not sum to 100\%, and 13 substantive comments were coded. Because the exercise uses a convenience sample, self-reported judgments, and a single representative report, it supports an analyst-in-the-loop pilot rather than a causal claim about productivity or decision quality.

\end{document}